\documentclass[letterpaper,twocolumn,american,showpacs,pr,aps,superscriptaddress,eqsecnum,nofootinbib]{revtex4}

\usepackage{amsfonts}%
\usepackage{amsmath}%
\usepackage{amssymb}%
\usepackage{graphicx}
\usepackage{color}
\usepackage{lipsum}
\usepackage{comment}
\usepackage{hyperref}

\newtheorem{theorem}{Theorem}

\newtheorem{corollary}[theorem]{Corollary}

\newtheorem{definition}[theorem]{Definition
}

\newenvironment{proof}[1][Proof]{\noindent\textbf{#1.} }{\ \rule{0.5em}{0.5em}}

\definecolor{nblue}{rgb}{0.2,0.2,0.7}
\definecolor{ngreen}{rgb}{0.2,0.6,0.2}
\definecolor{nred}{rgb}{0.7,0.2,0.2}
\definecolor{nblack}{rgb}{0,0,0}

\newcommand*{\bbR}{\mathbb{R}}

\begin{document}
\title{Extending Noether's theorem by quantifying the asymmetry of quantum states}
 

\author{Iman Marvian}
\affiliation{Perimeter Institute for Theoretical Physics, 31 Caroline St. N, Waterloo, \\
Ontario, Canada N2L 2Y5}
\affiliation{Institute for Quantum Computing, University of Waterloo, 200 University Ave. W, Waterloo, Ontario, Canada N2L 3G1}
\affiliation{Department of Physics and Astronomy, Center for Quantum Information Science and Technology, University of Southern California, Los Angeles, CA 90089}

\author{Robert W. Spekkens}
\affiliation{Perimeter Institute for Theoretical Physics, 31 Caroline St. N, Waterloo, \\
Ontario, Canada N2L 2Y5}

\date{\today}

\begin{abstract}

Noether's theorem is a fundamental result in physics stating that every symmetry of the dynamics implies a conservation law.  It is, however, deficient in several respects: (i) it is not applicable to dynamics wherein the system interacts with an environment, and (ii) even in the case where the system is isolated, if the quantum state is mixed then the Noether conservation laws do not capture \emph{all} of the consequences of the symmetries. To address these deficiencies, we introduce measures of the extent to which a quantum state breaks a symmetry. Such measures yield novel constraints on state transitions: for nonisolated systems, they cannot increase, while for isolated systems they are conserved. We demonstrate that the problem of finding nontrivial asymmetry measures can be solved using the tools of quantum information theory.  Applications include deriving model-independent bounds on the quantum noise in amplifiers and assessing quantum schemes for achieving high-precision metrology.

\end{abstract}

\maketitle


\section{Introduction}

Finding the consequences of symmetries for dynamics is a subject with broad
applications in physics, from high energy scattering
experiments, through control problems in mesoscopic physics, to issues in quantum cosmology.
In many cases, a complete solution of the dynamics is not possible either because it is too complex or because one lacks precise knowledge of all of the relevant parameters. In such cases, one can often still make nontrivial inferences by a consideration of the
symmetries. The most prominent example is the inference from dynamical symmetries to constants of the motion in closed
system dynamics.  For instance, from invariance of the laws of motion under translation in time, translation in space, and rotation, one can infer, respectively, the conservation of energy, linear momentum and angular momentum. 
This result has its origin in the work of Lagrange in classical mechanics~\cite{Goldstein}, but when the symmetries of interest are differentiable, the connection is established
by Noether's theorem \cite{Noether}. These days, physicists tend to use the term
``Noether's theorem'' to refer to the general
result, and we follow this convention here.  The theorem applies also in the quantum realm, where symmetries of the time evolution imply the existence of a set of observables all of whose moments are conserved~\cite{WeinbergQFT}.

A symmetric evolution is one that commutes with the action of the symmetry group~\cite{BRSreview}.
For instance, a rotationally-invariant dynamics is such that a rotation of the state prior to the dynamics has the same effect as doing so after the dynamics.
An \emph{asymmetry measure} 
quantifies how much the symmetry in question is broken by a given state.
More precisely, a function $f$ from states to real numbers is an asymmetry measure if the existence of symmetric dynamics taking $\rho$ to $\sigma$ implies $f(\rho)\ge f(\sigma)$~\cite{BRST06,GourSpekkens, Vaccaro, Toloui}.  A measure for rotational asymmetry, for instance, is a function over states that is non-increasing under rotationally-invariant dynamics.




For systems interacting with an environment (open-system dynamics),  where Noether's theorem does not apply, every asymmetry measure imposes a non-trivial constraint on what state transitions are possible under the symmetric dynamics, namely that the measure evaluated on the final state be no larger than that of the initial state.

For isolated systems (closed-system dynamics),  the existence of a symmetric unitary for some state transition implies the existence of a symmetric unitary for the reverse transition (namely, the adjoint of the unitary), hence each asymmetry measure is a conserved quantity under the symmetric dynamics.
We show that for transitions between mixed states, the conserved quantities one obtains in this way are independent of those prescribed by Noether's theorem. 
In this way, we find new conservation laws which are not captured by Noether's theorem. 

Our results also allow us to derive constraints on state transitions given \emph{discrete} symmetries of the dynamics, that is, symmetries associated with finite groups, where there are no generators of the group action and it is less obvious how to apply Noether's theorem.



\section{The inadequacy of Noether conservation laws for general dynamics}
How can we  find nontrivial asymmetry measures? In the case of rotational symmetry,  one might guess that the (expectation value of) components of angular momentum are good candidates.  For one,  a state with  nonzero angular momentum is necessarily non-invariant under some rotation.  For another, in closed system dynamics, any asymmetry measure must be a constant of the motion and angular momentum certainly satisfies this condition.  Nonetheless, it turns out that angular momentum is \emph{not} an asymmetry measure.
More generally, it turns out that none of the Noether conserved quantities, nor any functions
thereof, provide nontrivial measures of asymmetry.
To prove this, it is necessary to provide more precise definitions of the notions of symmetric operations and symmetric states.

\begin{figure}
 \center{   \includegraphics[width=7cm]{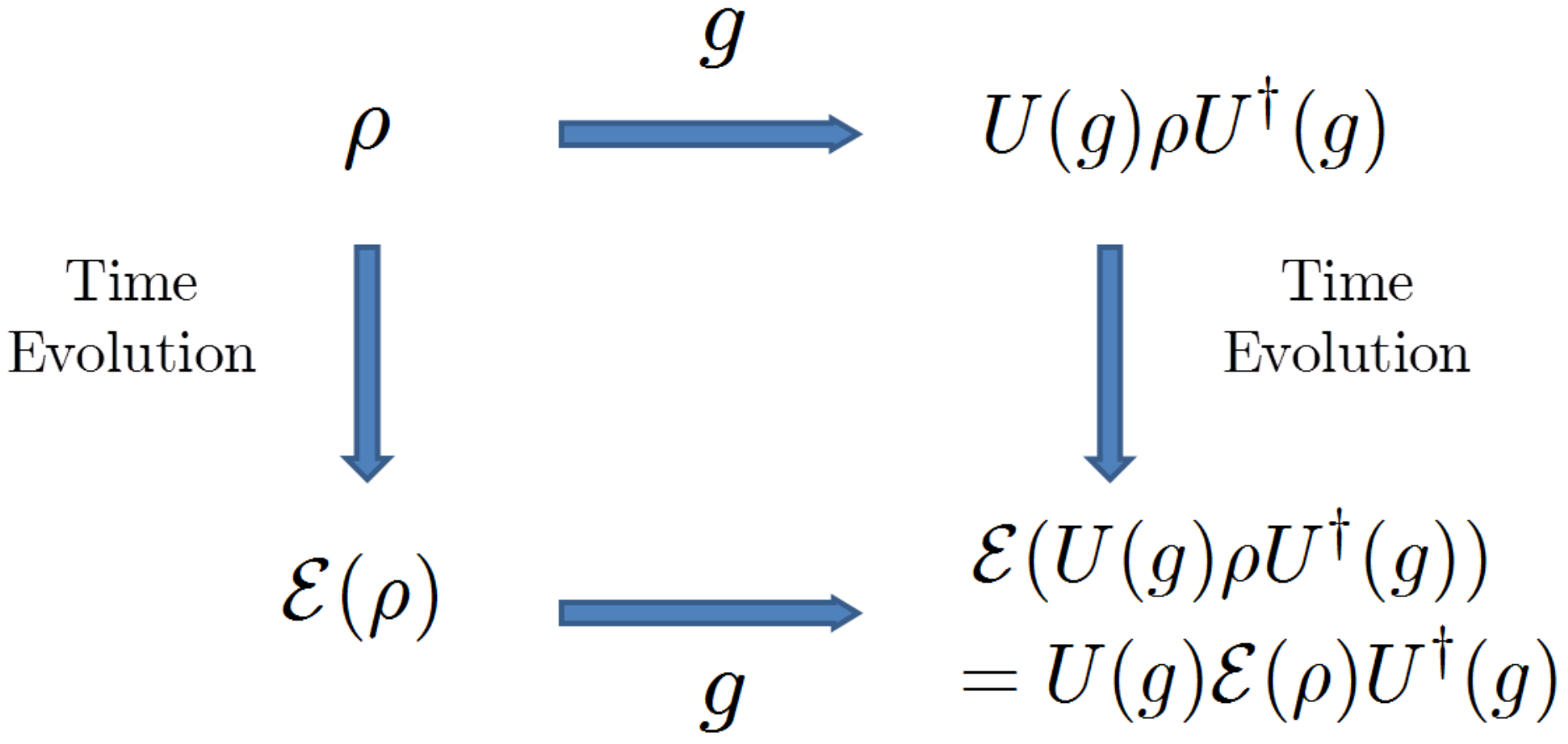}}
 \caption{\label{Fig:G-Equivalence}
A time evolution is called symmetric relative to a group $G$ if the map $\mathcal{E}$ describing the evolution commutes with the symmetry transformation associated with every group element $g\in G$.    }
\end{figure}

\begin{figure}
 \center{   \includegraphics[width=6cm]{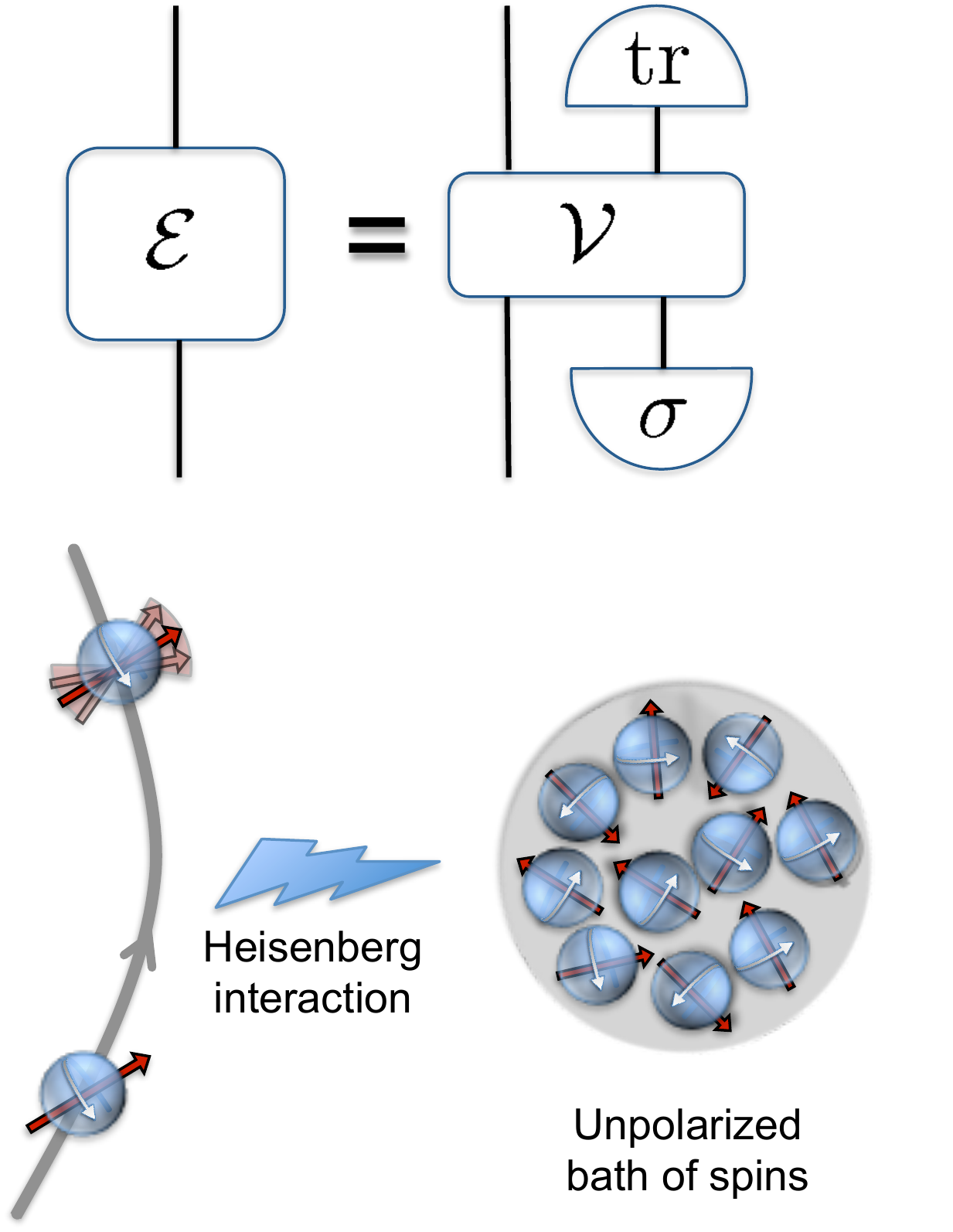}}
 \caption{\label{Fig:SymmetricOpenDynamics}
Symmetric open-system dynamics.  The top figure depicts such dynamics as a circuit. The system is coupled to an ancilla that is prepared in a symmetric state $\sigma$, i.e. $[\sigma,U(g)]=0\; \forall g\in G$, and the pair undergo a closed-system dynamics associated with a symmetric unitary map $\mathcal{V}(\cdot)\equiv V \cdot V^{\dag}$ where $V$ is a unitary operator satisfying the symmetry property $[V,U(g)]=0\; \forall g\in G$.  The overall dynamics for the system then corresponds to an open-systen dynamics associated with a completely-positive trace-preserving linear map $\mathcal{E}(\cdot)\equiv \textrm{tr}_a[V(\cdot \otimes \rho)V^{\dag}]$ that is symmetric, i.e. $\mathcal{E}(U(g) \cdot U^{\dag}(g))=U(g)\mathcal{E}(\cdot)U^{\dag}(g)\; \forall g\in G$. (here $\textrm{tr}_a$ denotes the trace operation on the ancilla).  Furthermore, every symmetric map $\mathcal{E}$ can be implemented in this fashion~\cite{KeylWerner}.
An example of such a dynamics is a spin-1/2 system interacting via a Heisenberg spin-spin interaction (which is rotationally-invariant) with a bath of spin-1/2 systems that are polarized in random directions, such that the overall state is unpolarized. This is depicted in the bottom figure.
}
\end{figure}

One specifies the symmetry
of interest by specifying an abstract group $G$ of transformations and the
appropriate  representation thereof. 
For a general symmetry described by a group $G$ the symmetry transformation corresponding to the group element $g$ is represented by the map $\rho\rightarrow \mathcal{U}_{g}(\rho)$ where $\mathcal{U}_g(\cdot)\equiv U(g)(\cdot)U^{\dag}(g)$ and $U(g)$ is a unitary operator. 
For instance, under rotation around  an axis $\hat{n}$ by angle $\theta$ the density operator of system will be transformed as $\rho\rightarrow e^{-i\theta\mathbf{J\cdot\hat{n}}} \rho e^{i\theta\mathbf{J\cdot\hat{n}}}$ where $\mathbf{J}=\left(  J_{x},J_{y}%
,J_{z}\right)  $ is the vector of angular momentum operators. 
A state $\rho$ does not break the symmetry $G$, or is symmetric relative to group $G$, if for all group elements $g\in G$ it holds that $\mathcal{U}_{g}(\rho)=\rho$.  
The concept of symmetry transformations and symmetric states can be naturally extended to the case of time evolutions. A time evolution $\mathcal{E}$ is  in general a linear transformation  from the space of density operators to itself, $\rho\rightarrow \mathcal{E}(\rho)$. We say the transformation $\mathcal{E}$ is symmetric relative to group $G$ if this transformation commutes with the symmetry transformation $\rho\rightarrow \mathcal{U}_{g}(\rho)$ for all group elements $g$ in group $G$ (See Fig.~\ref{Fig:G-Equivalence}).
Note that this definition of symmetric time evolution applies equally well to the cases of closed system dynamics, where the system does not interact with an environment, and open system dynamics (See Fig. \ref{Fig:SymmetricOpenDynamics}).

For a symmetry described by a Lie group $G$, Noether's theorem states that every 
generator $L$ of $G$  is conserved, or equivalently, the expectation value of any function of $L$ is conserved. Therefore, if under a unitary symmetric dynamics, state $\rho$ evolves to state $\sigma$, then all the implications of Noether's theorem can be summarized as:
\begin{equation} \label{Noether-condition}
\forall k \in \mathbb{N}:\ \  tr(\rho L^{k})=tr(\sigma L^{k})
\end{equation}
for every generator $L$ of $G$. 

We can now make precise the sense in which Noether conserved quantities yield no nontrivial measures of asymmetry:
For a symmetry corresponding to any compact Lie group $G$, any asymmetry measure $f$ that is a continuous function of only the Noether conserved quantities is trivial, that is, it takes the same value for all states. 

The proof is provided in the supplementary material, but we will here sketch the main idea.
 States that are asymmetric 
 must necessarily fail to commute with some generator of the group, say $L$, and consequently must have coherence between different eigenspaces of $L$ (for example, a state that is noninvariant under phase shifts necessarily has coherence between eigenspaces of the corresponding number operator). A nontrivial asymmetry measure must be able to detect such coherence.  If the state were known to be pure, then the presence of such coherences would be revealed by a nonzero variance over $L$.  However, an asymmetry measure is a function over \emph{all} states, and there exist \emph{mixed} states that have nontrivial variance over $L$ even though they have no coherence between the eigenspaces of $L$.  Hence the value of the second moment of $L$ has no information about the asymmetry properties of the state.  By the same logic, \emph{no} moment of $L$ has any information about the asymmetry properties of the state.

It follows that the problem of devising measures of asymmetry is nontrivial.  Nonetheless, it 
can be solved by taking an information-theoretic perspective on symmetric dynamics. 



\section{Building asymmetry measures: the power of the information-theoretic perspective}
\begin{figure}
\centering
\includegraphics[width=.45\textwidth,clip=true]{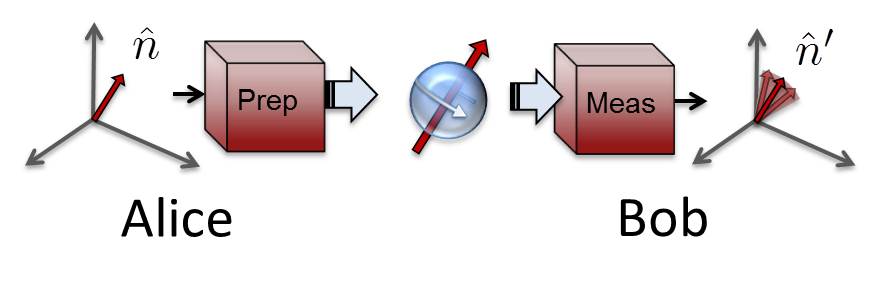}
\caption{\label{Fig_est1} {A schematic representation of a quantum communication protocol for sending information about direction: Alice chooses a direction $\hat{n}$ and prepares a spin-$j$ system in the coherent state in this direction, that is, she prepares the state $|j,j\rangle_{\hat{n}}$ where $\vec{L}\cdot\hat{n}|j,j\rangle_{\hat{n}}= \hbar j |j,j\rangle_{\hat{n}}$.  Then she sends this spin system to Bob. Bob performs a measurement on the system and obtains an estimate of $\hat{n}$, denoted $\hat{n}'$.  The uncertainty principle implies that there is a fundamental limit on the accuracy of Bob's estimate which is determined by $j$: the variance of the estimated angles are bounded by a constant factor times $1/j^{2}$.  Within the set of coherent states, therefore, the amount of directional information, and hence the rotational asymmetry, increases with $j$.}}
\end{figure}

Consider the problem of communicating information about a direction in space (See Fig.~\ref{Fig_est1}).   It is clear that to be able to succeed in this task,
one needs to use states which break the rotational symmetry. Furthermore, intuitively we expect that to transfer more directional information one needs to use states which are more asymmetric.
This suggests that one can quantify rotational asymmetry by the amount of information a state encodes about orientation.


To make this connection precise, we note that if $\rho \to \sigma$ by some symmetric dynamics then by definition this dynamics also takes every state in the group orbit of $\rho$ to the corresponding state in the group orbit of $\sigma$, that is, $\mathcal{U}_g(\rho) \to \mathcal{U}_g(\sigma)$ for all $g\in G$. 
The set of states $\{\mathcal{U}_g(\rho):g\in G \}$
can be understood as a \emph{quantum encoding} of the group element $g$,
and the dynamical evolution realizing $\mathcal{U}_g(\rho)\rightarrow \mathcal{U}_g(\sigma)$ for all $g\in G$ can be understood as a kind of data processing.  From the existence of such a data processing, it follows that the $\sigma$-based encoding must contain no more information about $g$ than the $\rho$-based encoding.


A measure of the information content of an encoding is a function from encodings to reals that is nonincreasing under data processing.  Specifically, $I$ is a measure of information if for any two different quantum encodings of a classical random variable $x \in X$, $\{ \rho_x:x\in X \}$ and $\{ \sigma_x:x\in X  \}$, the existence of a dynamical evolution that maps $\rho_x$ to $\sigma_x$ for all $x\in X$ implies that $I(\{ \rho_x:x\in X \}) \ge I(\{ \sigma_x:x\in X  \})$.  (In the context of information theory, 
the monotonicity of a measure of information is known as the data processing inequality.)
It follows that if we define a real function $f$ such that its value on a state is the measure of information $I$ of the group orbit of that state, that is, $f(\rho) \equiv I(\{\mathcal{U}_g(\rho):g\in G \})$, then $f$ is a measure of asymmetry.  The proof is simply that if $\rho$ is mapped to $\sigma$ by some symmetric dynamics, then for all $g\in G$, the state $\mathcal{U}_g(\rho)$ is mapped to $\mathcal{U}_g(\sigma)$ by that dynamics, and consequently $I(\{\mathcal{U}_g(\rho):g\in G \})\ge I(\{\mathcal{U}_g(\sigma):g\in G \})$, which implies $f(\rho)\ge f(\sigma)$.

Quantum information theorists have defined many measures of information and for each of these we can obtain a measure of asymmetry.  We mention a few that can be derived in this fashion (details of the derivation are provided in the supplementary material).

(i) Let $p(g)$ be an arbitrary probability density over the group manifold, and define the \emph{twirling operation weighted by $p(g)$} as $\mathcal{G}_{p} \equiv \int dg\ p(g)\ \mathcal{U}_{g}$.  Let $S(\rho) \equiv - \textrm{tr}(\rho \log \rho)$ be the von Neumann entropy.  
The function
\begin{equation} \label{eq:Holevo}
\Gamma_{p}(\rho)\equiv S\left(\mathcal{G}_{p}(\rho)\right)-S(\rho)
\end{equation} 
 is an asymmetry measure. We will refer to such a measure as a \emph{Holevo asymmetry measure}.   
The intuition behind it is as follows: 
if a state is close to symmetric, then it is close to invariant under rotations and mixing over all rotations does not change its entropy much, while if it is highly asymmetric, then under rotations it covers a broader manifold of states and hence mixing over all rotations increases the entropy significantly.

(ii) Let the matrix commutator of $A$ and $B$ be denoted by $[A,B]$ and the trace norm (or $\ell_1$-norm) by $\|A\|_1 \equiv \mathrm{tr} (\sqrt{A^{\dag}A})$. For any generator $L$ of the group action, the function 
\begin{equation} \label{eq:l1norm}
F_{L}(\rho)\equiv \|[\rho,L]\|_1
\end{equation} 
  is a measure of asymmetry. This measure formalizes the intuition that the asymmetry of a state can be quantified by the extent to which it fails to commute with the generators of the symmetry. 
  
(iii) For any generator $L$, and for $0<s<1$, the function 
\begin{equation} \label{eq:skew}
S_{L,s}(\rho)\equiv \text{tr}(\rho L^2)- \text{tr}(\rho^s L \rho^{1-s}L)
\end{equation} 
is a measure of asymmetry. 
This quantity was introduced by Wigner and Yanase with $s=1/2$ ~\cite{WignerYanaseDyson} and generalized by Dyson to arbitrary $s$.  While it has attracted much interest, its monotonicity under symmetric dynamics---and hence its interpretation as a measure of asymmetry---was not previously recognized. 

For all of these examples, if $\rho$ is a symmetric state then the asymmetry measure has value zero. 
Furthermore, for $F_L$ and $S_{L,s}$, if $\rho$ is a pure state then the measure reduces to the variance over $L$.  The measures $F_L$ and $S_{L,s}$ can both be understood as quantifying the ``coherent spread'' over the eigenvalues of $L$.  As discussed earlier, this is precisely what the variance over $L$ (or any function of $L$) could not do.  

\section{The inadequacy of Noether conservation laws for general closed-system dynamics}
Functions of the Noether conserved quantities cannot distinguish symmetric states from asymmetric states.
However, the examples we have provided thus far to establish this have relied on considering pairs of states that differ in their degree of purity (or entropy content), and if these are to be the input and output states of some dynamics, then the dynamics must be open.
We might hope, therefore, that although conservation of all Noether quantities is neither a necessary nor a sufficient condition for the possibility of a state transition
in \emph{open-system} dynamics, it is necessary and sufficient for \emph{closed-system} dynamics. 
Once again, however, we show that such hopes are not fulfilled.  




Consider a spin-1/2 system that also has some other independent degree of freedom which is invariant under rotation, denoted by the observable $Q$.  Let $|+\hat{n}\rangle,|-\hat{n}\rangle$ denote eigenstates of spin along the $\hat{n}$-axis, and let $|q_1\rangle, |q_2\rangle$ denote orthogonal eigenstates of $Q$.
Then define
\begin{equation}
\rho\equiv \tfrac{1}{2} |+\hat{z}\rangle \langle +\hat{z}| \otimes |q_1\rangle \langle q_1| 
+ \tfrac{1}{2} |-\hat{z}\rangle \langle -\hat{z}| \otimes |q_2\rangle \langle q_2|,
\end{equation}
 and
 \begin{equation}
 \sigma \equiv \tfrac{1}{2} |+\hat{x}\rangle \langle +\hat{x}| \otimes |q_1\rangle \langle q_1| 
+ \tfrac{1}{2} |-\hat{x}\rangle \langle -\hat{x}| \otimes |q_2\rangle \langle q_2|.
 \end{equation}
We can easily check that:
(i) the state transition $\rho \to \sigma$ is impossible by rotationally-symmetric closed-system dynamics (but possible by dynamics that break rotational symmetry, namely, a rotation around $\hat{y}$ by $\pi/2$, so that the transition would \emph{not} be forbidden were it not for considerations of symmetry)
(ii) All constraints implied by Noether's theorem hold, i.e.,  the conditions of Eq.~(\ref{Noether-condition}) for generators of rotations are satisfied, and therefore Noether's theorem does not forbid this transition.


Condition (ii) is straightforward to verify. The truth of (i) can be made intuitive by noting that $\rho$ is symmetric under rotations about the $\hat{z}$-axis while $\sigma$ is not, such that a transition from $\rho$ to $\sigma$ is symmetry-breaking and hence impossible by rotationally-symmetric dynamics. One can derive this same conclusion using asymmetry measures.  Consider the Holevo asymmetry measure
$\Gamma_{p}$ for the probability density  $p: \textrm{SO(3)} \to \bbR_+$ that is uniform over all rotations around $\hat{z}$ and vanishing for all other rotations. We find that
$ \Gamma_{p}(\rho)=0$ and $\Gamma_{p}(\sigma)=1$,
that is, $\Gamma_{p}$ \emph{increases} in this state transition, thereby demonstrating that it cannot be achieved by rotationally-symmetric dynamics.
We have shown that  constants of the motion derived from asymmetry measures can capture restrictions on the dynamics that are not captured by Noether's theorem.

\section{The adequacy of Noether conservation laws for closed-system dynamics of pure states}
One final special case remains to be considered.  Might it be that conservation of the Noether conserved quantities are the necessary and sufficient conditions for the possibility of state interconversion under symmetric closed-system dynamics \emph{when the states are pure}?  In this case, the answer is yes.

In~\cite{MarvianSpekkens2013}, it was shown that the asymmetry properties of a pure state $|\psi \rangle$ are completely determined by the complex function $\langle\psi|U(g)|\psi\rangle$ over the group manifold, called the characteristic function.  
Equality of characteristic functions is the necessary and sufficient condition for two pure states to be reversibly interconvertible under symmetric unitary dynamics. Expanding $U(g)$ in a power series over the generators, one deduces that for connected compact Lie groups, such as the group of rotations,
equality of all moments of the generators is equivalent to equality of the characteristic functions.  
Consequently, for pure states undergoing reversible dynamics with such a symmetry, Noether's theorem, i.e. Eq.~(\ref{Noether-condition}), captures all of the consequences of the symmetry.


In practice, we are always faced with some loss of information under any quantum dynamics, due to the ubiquity of decoherence, and there is always some noise in our preparation of the initial state.  Therefore, reversible dynamics of pure states is an idealization that is never achieved in practice, and as soon as one departs from it, Noether's theorem is  inadequate for describing the consequences of symmetry.




\section{Applications of measures of asymmetry}
Phase-insensitive quantum amplifiers are examples of open system dynamics that have a symmetry property.  Consequently, from the nonincrease of measures of asymmetry, we can derive bounds on their performance.
The purpose of a quantum amplifier is to increase the expectation value of some observable, such as the number operator for optical fields~\cite{Caves}.
In most studies, constraints on amplification are obtained for specific physical models of the amplifier. Furthermore, the analysis is typically done separately for linear and nonlinear amplifiers as well as for deterministic and nondeterministic amplifiers~\cite{Caves2}.  By contrast, the constraints that can be found with our techniques follow from assumptions of symmetry alone.   For instance, in optics they follow from the fact that the amplifier is phase-insensitive. They are therefore model-independent and can be applied whether the amplifier is linear or nonlinear, deterministic or nondeterministic.

Here is an example of such a constraint, arising from the Holevo measure of asymmetry.  If $\rho$ is mapped to $\sigma$ by a symmetric amplifier, then
\begin{equation}\label{amplifier}
S(\sigma)-S(\rho) \ge S\left(\mathcal{G}_{p}(\sigma\right) )-S\left(\mathcal{G}_{p}(\rho)\right),
\end{equation}
which asserts that the change in entropy under the transition has a nontrivial lower bound.

For instance, suppose that $\rho$ is a state of a spin-1/2 system, while $\sigma$ is a state of a spin-$j$ system for $j \gg 1/2$. Suppose further that
the probability density $p$ in $\mathcal{G}_p$ is chosen to be the uniform measure over the symmetry group (which in this problem is $SO(3)$), so
that  $S\left(\mathcal{G}_{p}(\rho)\right)=1$ while $S\left(\mathcal{G}_{p}(\sigma\right) )$ is large (logarithmic in $j$).  The inequality then implies that $S(\rho)-S(\sigma)$ must be large.

\begin{figure}
\centering
\includegraphics[width=.45\textwidth,clip=true]{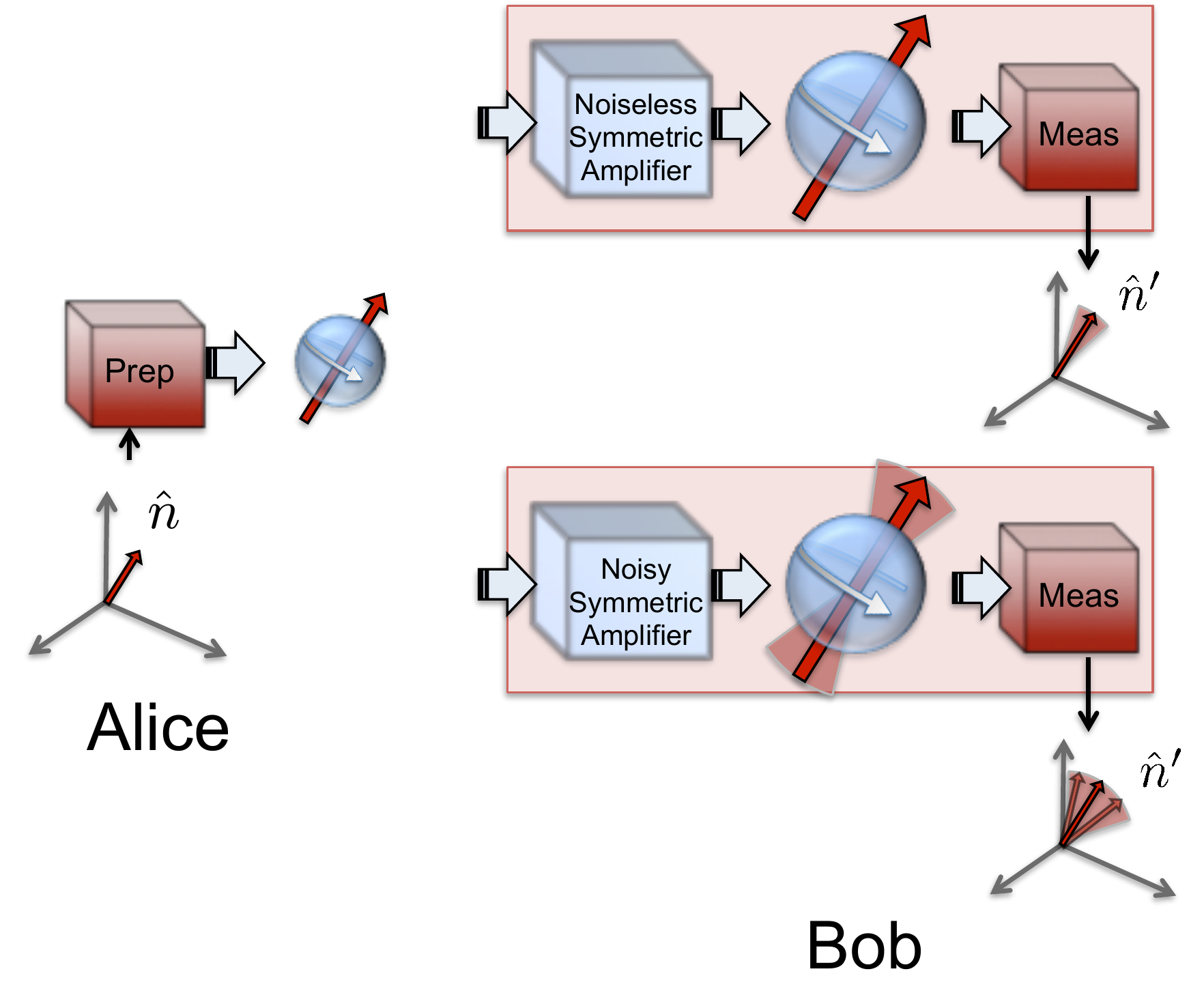}
\caption{\label{Fig_est2}
An example of how one can derive constraints on state transitions in symmetric open-system dynamics from considerations of how much information the states encode about the symmetry group. Suppose that there were a rotationally-symmetric open-system dynamics which transforms the coherent state aligned in the $\hat{n}$-direction, $|j,j\rangle_{\hat{n}}$, to the coherent state $|j',j'\rangle_{\hat{n}}$ where $j'>j$, i.e., a rotationally-symmetric noiseless amplifier.  If this were possible  then in the communication protocol described in Fig.~\ref{Fig_est1}, Bob could first apply this dynamics and then estimate the direction $\hat{n}$ using the state $|j',j'\rangle_{\hat{n}}$. Such a strategy would yield an estimate with variance proportional to $1/j'^{2}$, which is better than what is allowed by the fundamental quantum limit. So we conclude that the state transition is not possible by rotationally-symmetric dynamics.
On the other hand, for a similar rotationally-symmetric dynamics which adds noise to the state, such that $|j,j\rangle_{\hat{n}}$ is not mapped to $|j',j'\rangle_{\hat{n}}$ but to a mixed spin-$j'$ state, as long as the noise is large enough to ensure that Bob can only estimate $\hat{n}$ with a variance proportional to $1/j^{2}$, considerations of information content do not rule out the transition.  Indeed, as we show in the text, one can always increase the expectation value of the angular momentum if there is a compensating increase in the entropy.}
\end{figure}

This demonstrates 
that in the case of rotationally-symmetric open-system dynamics, an increase in the value of the angular momentum along some axis is not prohibited as long as entropy increases.  This ensures that the distinguishability of states at the output is not more than the distinguishability of states at the input, so that the information content has not increased (See Fig.~\ref{Fig_est2}). 

A second application of measures of asymmetry is to quantify quantum coherence~\cite{Aberg}. Coherence is at the heart of many distinctly quantum phenomena from interference of individual quanta to superconductivity and superfluidity.  On the practical side, coherence is the property of quantum states that is critical for quantum phase estimation:
a coherent superposition of number eigenstates, such as $\frac{1}{\sqrt{2}}(|0\rangle+|n\rangle)$, is sensitive to phase shifts, while an incoherent mixture, such as $\frac{1}{2}(|0\rangle\langle 0|+|n\rangle\langle n|)$, is not.
Phase-shifts, however, are symmetry transformations. Therefore, states with coherence are precisely those that are asymmetric relative to the group of phase shifts. It follows that we can define a measure of coherence as any function that is nonincreasing under phase-insensitive time evolutions (See \cite{MarvianSpekkensModes} for more details). It follows that  the measures of asymmetry proposed here---Eqs.~\eqref{eq:Holevo},  \eqref{eq:l1norm} and \eqref{eq:skew}---can be used as measures of coherence relative to the eigenspaces of the generator $L$ (i.e. as the ``coherent spread'').

Finally, measures of asymmetry are important because asymmetry is the resource that powers quantum metrology.\footnote{This contrasts with the more standard view that the relevant resource is entanglement.   Note, however, that measures of entanglement can sometimes provide bounds on measures of asymmetry.}   Metrology involves estimating a symmetry transformation.  In the most general case, the set of symmetry transformations form a non-Abelian group, such as the group of rotations, but in the most common example, it is the Abelian group of phase shifts (in which case asymmetry corresponds to coherence).  We focus on this example to illustrate the idea. 

To estimate an unknown phase shift $\phi$ of an optical mode, one prepares the mode in the state $\rho$, subjects it to the unknown phase shift, leaving it in the state $e^{i\phi N}\rho e^{-i\phi N}$, where  $N$ is the number operator, and finally one measures it. 
The usefulness of a particular state $\rho$ can be quantified by any measure of the information content of the ensemble $\{ e^{i \phi N}\rho e^{-i \phi N} : \phi \in [0,2\pi)\}$, but, as shown above, every such measure is a measure of the asymmetry of $\rho$ relative to phase shifts. 
The figure of merit for a metrology task is therefore a measure of asymmetry and dictates the optimal $\rho$.   
Suppose, for instance, that one seeks an unbiased estimator $\hat\phi$ of $\phi$ and the figure of merit is the variance in $\hat\phi$, denoted $\textrm{Var}(\hat\phi)$.  It has been shown in~\cite{Luo} that for a state $\rho$, we have $\textrm{Var}(\hat\phi) \le  1/4 S_{N,\frac{1}{2}}(\rho)$ (a quantum generalization of the Cramer-Rao bound) where $S_{N,\frac{1}{2}}(\rho)$ is the Wigner-Araki-Dyson skew information of order $s=1/2$, defined in Eq.~\eqref{eq:skew}.  So it is the latter measure of asymmetry that is relevant in this case.  

\section{Discussion}
Our analysis prompts the question: what are the necessary and sufficient conditions on states $\rho$ and $\sigma$ (not both pure) for it to be possible to map $\rho$ to $\sigma$ under symmetric dynamics?  Such conditions would capture \emph{all} of the consequences of the symmetry of the dynamics.   The question remains open but our results suggest that adopting an information-theoretic perspective may be the most expedient path to a solution.

\color{black}

\color{black}

\section{Acknowledgements}
Research at Perimeter Institute is supported in part by the Government of Canada through NSERC and by the Province of
Ontario through MRI. IM acknowledges support from NSERC, a Mike and Ophelia Lazaridis fellowship, and ARO MURI grant  W911NF-11-1-0268.

\section{Supplementary material}


\subsection{Triviality of asymmetry measures based on Noether conserved quantities}

In this section, we prove that for symmetries corresponding to compact Lie groups, functions of Noether conserved quantities yield only trivial measures of asymmetry.  In the case of finite groups, there are no generators of the group action, and therefore we cannot generate Noether conserved quantities in the standard way.  Nonetheless, we can show that a similar result holds in this case as well.

We phrase our general result (which applies to both compact Lie groups and finite groups) in terms of \emph{characteristic functions}. Recall that the characteristic function of a state $\rho$ is the expectation value of the unitary representation of the group, $\text{tr}(\rho U(g))$, which is a complex function over the group manifold.  Then we can prove the following theorem about measures of asymmetry.
\color{black}
\begin{theorem} \label{thm:generalresult}
Let $f$ be an asymmetry measure for a finite or compact Lie group $G$ with unitary representation $U$.  Assume that $f$ is a function of the characteristic function of the state alone, i.e., $f(\rho)=\mathcal{F}[\text{tr}(\rho U(g))]$ for some functional $\mathcal{F}:\mathbb{C}(G)\rightarrow \mathbb{R}$. Furthermore, in the case of compact Lie groups, assume that $f$ is continuous. Then the monotone $f$ is a constant function, i.e., $f(\rho)$ is independent of $\rho$.
\end{theorem}

In the case of compact Lie groups, the characteristic function of a state $\rho$ uniquely specifies all the moments $\text{tr}{(\rho L^{k})}$ for all generators $L$ of the symmetry; this can be seen by considering the Taylor expansion of $U(g)$ around the identity. Therefore, the theorem implies that if a continuous measure of asymmetry can be expressed entirely in terms of the Noether conserved quantities alone, i.e. in terms of moments of the form $\textrm{tr}(\rho L^k)$, then it should be a constant function independent of $\rho$.

We now present the proof of theorem~\ref{thm:generalresult}.

\begin{proof}
We first present the proof for the case of finite groups and then we explain how the result can be generalized to the case of compact Lie groups as well. 



Suppose $\mathcal{H}_*$ is the Hilbert space of a physical system on which the unitary representation $g\rightarrow U(g)$ of the symmetry group $G$ acts as the left regular representation, i.e., $\mathcal{H}_*$ has an orthonormal basis denoted by $\{|g\rangle:\ g\in G\}$ such that
\begin{equation}
\forall g,h\in G:\ U(g)|h\rangle=|gh\rangle.
\end{equation}
It turns out that on this space we can  define another representation of $G$, called the \emph{right regular representation} denoted by $g\rightarrow V_{R}(g)$, such that
\begin{equation}
\forall g,h\in G: V_{R}(g)|h\rangle=|hg^{-1}\rangle
\end{equation}
Then one can easily see that these two representations of $G$ on $\mathcal{H}_*$ commute, i.e.
\begin{equation}\label{LR-commuting}
\forall g,h\in G:\ [V_{R}(h),U(g)]=0
\end{equation}
Define $\mathcal{G}$ to be the quantum operation that averages over all symmetry transformations, $\mathcal{G} \equiv \sum_{g\in G} \mathcal{U}_{g}$. Let $e\in G$ be the identity element of the group.  We have 
\begin{align*}
&\text{tr}\left(U(g)\mathcal{G}(|e\rangle\langle e|)\right) \\
&=\frac{1}{|G|}\sum_{s\in G} \text{tr}\left(U(g)U(s)|e\rangle\langle e|U^{\dag}(s)\right)\\ 
&=\frac{1}{|G|}\sum_{s\in G} \text{tr}\left(U(g)V_{R}(s^{-1})|e\rangle\langle e|V_{R}^{\dag}(s^{-1})\right)\\ 
&=\text{tr}\left(U(g)|e\rangle 
\langle e|\right),
\end{align*}
where to get the second equality we have used the fact $\forall h\in G:\ \mathcal{G}(X)=\mathcal{G}(U(h)X U^{\dag}(h))$ and to get the last equality we have used the fact that the two representations commute, Eq.~(\ref{LR-commuting}). So the characteristic function of the state $|e\rangle \langle e|$ is equal to the characteristic function of the state $\mathcal{G}(|e\rangle\langle e|)$. This means that for any measure $f$ whose value for a given state depends only on the characteristic function of that state, we have
\begin{equation}
f\left(|e\rangle\langle e|\right)= f\left(\mathcal{G}(|e\rangle\langle e|)\right)
\end{equation}
As we have seen before, however, for any asymmetry measure, the value of the measure is the same for all symmetric states and furthermore this value is the minimum value of that function over all states.  Given that $\mathcal{G}(|e\rangle\langle e|)$ is a symmetric state, it follows that
\begin{equation}\label{eq-min-monotone}
f\left(|e\rangle\langle e|\right)= \min_{\sigma} f\left(\sigma\right).
\end{equation}


Now consider an arbitrary state $\rho$ on an a Hilbert space $\mathcal{H}$ where the projective unitary representation of the symmetry group $G$ is  $g\rightarrow T(g)$. Then, one can easily show that there exists symmetric quantum channels which map the state $|e\rangle\langle e|$ on $\mathcal{H}_*$ to the state $\rho$ on $\mathcal{H}$. One such channel is described by the map
\begin{equation}
\mathcal{E}_{\rho}(X)\equiv \frac{1}{|G|}\sum_{g\in G} \text{tr}\left(|g\rangle\langle g|X\right) T(g) \rho T^{\dag}(g)
\end{equation}
But the fact  that $\mathcal{E}_{\rho}$ is symmetric together with the fact that $f$ is an asymmetry measure implies that for any state $\rho$ it holds that
 \begin{equation}
 f\left(\rho\right)=f\left(\mathcal{E}_{\rho}(|e\rangle\langle e|)\right) \le f\left(|e\rangle\langle e|\right)
\end{equation}
This together with Eq.~(\ref{eq-min-monotone}) implies that  for an arbitrary state $\rho$,
 \begin{equation}
 f\left(\rho\right)= \min_{\sigma} f\left(\sigma \right),
\end{equation} 
and so the asymmetry measure $f$ is constant over all states. This completes the proof for the case of finite groups. 

In the following, we prove that making the extra assumption that  the asymmetry measure is  also continuous, this result can be extended to the case of compact Lie groups. Note that in this case the regular representation of the group is not finite dimensional.  Nonetheless, as was noted in  \cite{Kitaev-Mayers-Preskill} and later in \cite{BRSreview}, there still exists a sequence of finite dimensional  spaces, $\{ \mathcal{H}_{d} \}$, where $d$ is the maximum dimension of irreducible representation (irrep) supported on each space, and for each $\mathcal{H}_d$ there is an over-complete basis $\{|g\rangle:g\in G\}$, such that the unitary representation $g\rightarrow U(g)$ of the symmetry group $G$ acts as\footnote{ Let $\mathcal{H}_{d}=\bigoplus_{\mu:\\ d_{\mu}\le d} \mathcal{M}_{\mu}\otimes \mathcal{N}_{\mu}$ where the summation is over all irreps of $G$ whose dimension $d_{\mu}$ is less than or equal to $d$, and  $\mathcal{M}_{\mu}$ is the subsystem on which the symmetry $G$ acts like its irrep $\mu$ and $\mathcal{N}_{\mu}$ is the multiplicity subsystem with dimension equal to $d_{\mu}$. Define $|e_{*}\rangle=c \sum_{\mu} \sqrt{d_{\mu}}\sum_{i=1}^{d_{\mu}} |\mu,i\rangle\otimes|\tilde{i}\rangle $  where $c$ is a normalization factor, $\{|\mu,i\rangle: i=1\cdots d_{\mu}\}$ is an orthonormal basis for subsystem $\mathcal{M}_{\mu}$, and  $\{|\tilde{i}\rangle: i=1\cdots d_{\mu}\}$ is an orthonormal basis for $\mathcal{N}_{\mu}$.  The properties assumed for $\{|g\rangle=U(g)|e\rangle:\ g\in G\}$ in the proof hold if this set of states is generated from a fiducial state $|e\rangle$ of the form of  $|e_*\rangle$.} 
\begin{equation}
\forall g,h\in G:\ U(g)|h\rangle=|gh\rangle.
\end{equation}

Furthermore, as is discussed in  \cite{Kitaev-Mayers-Preskill} and \cite{BRSreview}, 
for a given pair of distinct group elements, $g_{1}\neq g_{2}$, one can make the inner product $\langle g_{2}|g_{1}\rangle$ arbitrarily close to zero
in the limit of large $d$.
In this limit, the state $|e\rangle \langle e|$ has the \emph{maximal asymmetry} in the sense that for any given state $\rho$ on an aribrary Hilbert space $\mathcal{H}$, 
there exists a symmetric channel $\mathcal{E}_{\rho}$ such that 
\begin{equation}\label{Eq-monotone-limit}
\lim_{d\rightarrow \infty}\mathcal{E}_{\rho}(|e\rangle\langle e|)\rightarrow \rho. 
\end{equation}
The symmetric channel  $\mathcal{E}_{\rho}$  can be defined in a manner similar to how it was defined for the case of finite  groups, namely,
\begin{equation}
\mathcal{E}_{\rho}(X)\equiv \int dg\  \text{tr}\left(|g\rangle\langle g|X\right) T(g) \rho T^{\dag}(g).
\end{equation}

Now, similarly to the case of finite groups, we can also define another representation of $G$ on $\mathcal{H}_{d}$, denoted $g\rightarrow V_{R}(g)$, such that
\begin{equation}
\forall g,h\in G: V_{R}(g)|h\rangle=|hg^{-1}\rangle
\end{equation}
Then one can easily see that these two representations of $G$ on $\mathcal{H}_d$ commute, 
\begin{equation}\label{LR-commuting2}
\forall g,h\in G:\ [V_{R}(h),U(g)]=0
\end{equation}
Therefore, using the same argument that we used for the case of finite groups, we can prove that  for any $h\in G$ it holds that $\text{tr}\left(|h\rangle\langle h| U(g)\right)=\text{tr}\left(\mathcal{G}(|e\rangle\langle e|) U(g)\right)$ where $e$ is the identity element of the group $G$. Therefore, for any measure $f$ whose value for a given state depends only on the characteristic function of that state, it holds that
\begin{equation}
f\left(|e\rangle\langle e|\right)= \min_{\sigma} f\left(\sigma\right).
\end{equation}

Furthermore, because $f$ is an asymmetry measure and because $\mathcal{E}_{\rho}$ is a symmetric channel, we have
\begin{equation}
f\left(\mathcal{E}_{\rho}(|e\rangle\langle e|)\right) \le f(|e\rangle\langle e|).
\end{equation}
The above two equations imply that 
\begin{equation}\label{Eq-mon-Lie-upper}
f\left(\mathcal{E}_{\rho}(|e\rangle\langle e|)\right)=\min_{\sigma} f\left(\sigma\right).
\end{equation}
On the other hand, given that $f$ is assumed to be continuous, Eq. (\ref{Eq-monotone-limit}) implies that 
\begin{equation}
\lim_{d\rightarrow \infty}f\left(\mathcal{E}_{\rho}(|e\rangle\langle e|)\right)\rightarrow f(\rho).
\end{equation}
This, together with Eq. (\ref{Eq-mon-Lie-upper}), proves that for any arbitrary state $\rho$ in an arbitrary finite dimensional space $\mathcal{H}$, 
it holds that $f(\rho)=\min_{\sigma} f\left(\sigma\right)$.  Therefore, we conclude that 
in the case of compact Lie groups any continuous asymmetry measure which only depends on the characteristic function of the state is a constant function. This completes the proof. 
\end{proof}

\subsection{Some nontrivial families of asymmetry measures}

We now apply the recipe described in the article to generate a few interesting measures of asymmetry from measures of information.

Our first example makes use of a family of information measures that are based on the \emph{Holevo quantity}~\cite{Nielsen}. For a set of states $\{ \rho_x:x\in X\}$, and a probability distribution $p_x$ over $X$, the Holevo quantity is defined as 
$$H_{p}(\{ \rho_x:x\in X\})\equiv S\left(\sum_{x\in X} p_x \rho_x\right) - \sum_{x\in X} p_x S(\rho_x)$$ 
where $S(\rho)\equiv -\text{tr}(\rho\log \rho)$ is the von Neumann entropy of the state $\rho$
(if $x$ is a continuous variable, and $p(x)$ is a probability density, we simply replace sums by integrals). It is well-known that this quantity is non-increasing under information processing, i.e. it is an information monotone \cite{Nielsen}.
 This yields a family of asymmetry measures, one for every probability density $p(g)$ over the group manifold (probability distribution for the case of a finite group), namely,
\begin{equation}\label{Def-asymm-Holevo}
\Gamma_{p}(\rho)\equiv S\left(\mathcal{G}_{p}(\rho)\right)-S(\rho)
\end{equation}
where 
$$\mathcal{G}_{p} \equiv \int dg\ p(g)\ \mathcal{U}_{g}$$
 is the superoperator that performs a $p$-weighted average over the group action (which is sometimes called the \emph{twirling operation weighted by $p(g)$}). We call measures of this form \emph{Holevo asymmetry measures}. 
 
Note that for any symmetric state $\rho$ and any arbitrary probability distribution $p(g)$, $\Gamma_{p}(\rho)=0$. Also, note that for any  probability distribution $p(g)$ which is nonzero for all $G$, and for any state which breaks the symmetry, $\Gamma_{p}(\rho)\neq 0$.
For the special case of a uniform weighting, this measure has been previously proposed in Ref.~\cite{Vaccaro} and proven to be monotonic under symmetric operations using a different type of argument.

A particularly simple subclass of measures of the information content of a set of states are measures that consider the distinguishability of just a single pair of states within the set.  
For a pair of states $\rho_1$ and $\rho_2$, a measure of their distinguishability is defined to be a function $D$ from pairs of states to the reals, such that for any quantum channel $\mathcal{E}$, we have
\begin{equation}\label{distinguishabilitymeasure}
D\left(\mathcal{E}(\rho_{1}),\mathcal{E}(\rho_{2})\right) \le D\left(\rho_{1},\rho_{2}\right).
\end{equation}

Specializing to the case of interest here, where the classical variable $x \in X$ is a variable $g \in G$ (which ranges over the elements of the relevant group), we focus on the distinguishability of two elements in the group orbit of $\rho$.  Without loss of generality, we can choose the pair to be $\rho$ and  $\mathcal{U}_{g}(\rho)$ for some $g\in G$.  

We now introduce two more families of asymmetry measures based on this subclass of measures of information.  


Consider a Lie group $G$. Take the distinguishability measure to be the trace distance, 
\begin{equation}
D_{\text{tr}}(\rho_1,\rho_2)\equiv \|\rho_1 - \rho_2\|_1,
\end{equation}
where $\|X\|_1\equiv \text{tr}\left(\sqrt{X X^{\dag}}\right)$
is the trace norm (or $\ell_1$-norm).  It is well-known that the trace distance satisfies Eq.~\eqref{distinguishabilitymeasure} and hence constitutes a measure of the distinguishability of a pair of states~\cite{Helstrom}. 
Therefore, we can define an asymmetry monotone using this distinguishability measure, namely, $F_g(\rho)\equiv \| \rho - \mathcal{U}_g(\rho) \|_1$.

Now recall that for a Lie group $G$, we can consider a pair of states $\rho$ and $\mathcal{U}_g(\rho)$ where $g$ is infinitessimally close to the identity element. Specifically, we can always write $U(g)= e^{i\theta L}$ for some generator $L$ and phase $\theta$, and in the limit where $\theta \to 0$, we have
\begin{equation}
\|\rho-\mathcal{U}_g(\rho)\|_1 \simeq\theta \|[\rho,L]\|_1+\mathcal{O}(\theta^{2}).
\end{equation}
So we conclude that for any generator $L$ of the Lie group the function
\begin{equation}
F_{L}(\rho)\equiv\|[\rho,L]\|_1
\end{equation}
is an asymmetry measure. 

Any state $\rho$ that is symmetric (i.e. invariant under the group action) necessarily commutes with all the generators, so for such states, $F_L(\rho)=0$ for all $L$. 
Also, any state $\rho$ that is invariant only under some subgroup of $G$ has $F_L(\rho)=0$ for those $L$ that are generators of this subgroup.  

A state $\rho$ can only be asymmetric relative to a subgroup of $G$ associated with a generator $L$ if it has some coherence over the eigenspaces of $L$, that is, if  $[\rho,L]\ne 0$.  Therefore, in retrospect one would naturally expect that \emph{some} operator norm of the commutator $[\rho,L]$ should be a measure of asymmetry.  This intuition does not, however, tell us \emph{which} operator norm to use.  Our result shows that it is the trace norm that does the job.

$F_{L}$ also reduces to a simple expression for pure states: it 
is proportional to the square root of the variance of the observable $L$, that is,
\begin{equation}
F_{L}(|\psi\rangle\langle\psi|)=2 \left( \langle\psi| L^{2}|\psi \rangle-\langle\psi| L|\psi \rangle^{2} \right)^{1/2}.
\end{equation}

Given that a superposition over the eigenspaces of $L$ that is totally incoherent (i.e. a mixture over the eigenspaces) has vanishing asymmetry according to this measure, while a superposition over these eigenspaces that is totally coherent has asymmetry that depends only on the variance over $L$, this asymmetry measure seems to succeed in quantifying the amount of variance over $L$ that is coherent, which one might call ``coherent spread'' over the eigenspaces of $L$. 



We turn to our third and final example of a family of asymmetry measures.  We take as our measure of distinguishability the \emph{relative Renyi entropy} of order $s$, defined as
\begin{equation} \label{def}
D_{s}(\rho_{1},\rho_{2})\equiv \frac{1}{s-1}\log\left(\text{tr}(\rho_{1}^{s}\rho_{2}^{1-s})\right),
\end{equation}
For $s\in (0,1)\cup (1,\infty)$, it is well-known that $D_s$ satisfies Eq.~\eqref{distinguishabilitymeasure} and is therefore a valid measure of distinguishability~\cite{Hayashi}.  It follows that $F_{s,g}(\rho)\equiv D_s ( \rho, \mathcal{U}_g(\rho) )$ is an asymmetry measure. 
As in the previous example, by considering $D_s ( \rho, \mathcal{U}_g(\rho) )$ for group elements which are infinitessimally close to the identity element, we can derive a measure of asymmetry  for any arbitrary generator $L$ of the group. Using this argument, we can show that for any $s\in (0,1)\cup (1,\infty)$ and any generator $L$,
\begin{equation}
S_{L,s}(\rho)\equiv \text{tr}(\rho L^{2})-\text{tr}(\rho^{s}L \rho^{(1-s)}L)
\end{equation}
is a measure of asymmetry.


This family of measures has been previously studied under the name of the Wigner-Yanase-Dyson skew information~\cite{WignerYanaseDyson}, but their status as measures of asymmetry had not been recognized. 
\color{black}
If $\rho$ is a symmetric state, then it commutes with $L$, and $S_{L,s}(\rho)=0$.  We also find this measure to be zero when $\rho$ is not symmetric, but has some subgroup of $G$ as a symmetry and $L$ is a generator of that subgroup.    

For pure states, the Wigner-Yanase-Dyson skew information reduces to the variance of the observable $L$, that is, 
\begin{equation}
S_{L,s}(|\psi\rangle\langle\psi|)=  \langle\psi| L^{2}|\psi \rangle-\langle\psi| L|\psi \rangle^{2}.
\end{equation}

Again, we see that a mixture over the eigenspaces of $L$ has vanishing $S_{L,s}$, while a coherent superposition over these eigenspaces has $S_{L,s}$ equal to the variance over $L$.  Consequently, this asymmetry measure, like $F_L$, in some sense quantifies the coherent spread over the eigenspaces of $L$.





\begin{thebibliography}{000}


\bibitem {Goldstein}H. Goldstein, \textit{Classical Mechanics} (2nd ed.). Reading, MA: Addison-Wesley Publishing. pp. 588-596 (1980).
\bibitem{Noether} E. Noether, Nachr. D. Kšnig. Gesellsch. D. Wiss. Zu G\"{o}ttingen, Math-phys. Klasse, 235 (1918).
\bibitem{WeinbergQFT}  S. Weinberg, \textit{The Quantum Theory of Fields, Volume 1: Foundations.}  Cambridge University Press (2005). 

\bibitem{BRSreview} S. D. Bartlett, T. Rudolph, and R. W. Spekkens, Rev.
Mod. Phys. \textbf{79}, 555 (2007).

\bibitem{BRST06} S. D. Bartlett, T. Rudolph, R.W. Spekkens and P. S. Turner, New J. Phys. \textbf{8}, 58 (2006).

\bibitem{GourSpekkens}  G. Gour and R. W. Spekkens, New J. Phys. \textbf{10}, 033023  (2008).

\bibitem {Vaccaro}J.~A.~Vaccaro, F.~Anselmi, H.~M.~Wiseman, and K.~Jacobs, Phys. Rev. A \textbf{77}, 032114 (2008).

\bibitem{Toloui} B.~Toloui, G.~Gour, B.~C.~Sanders, Phys. Rev. A {\bf 84}, 022322 (2011).


\bibitem{KeylWerner} M. Keyl and R. F. Werner, J. Math. Phys. {\bf 40}, 3283 (1999).

\bibitem{WignerYanaseDyson} E. P. Wigner and M. M. Yanase, Proc. Nat. Acad. Sci. USA \textbf{49}, 910 (1963).

\bibitem{MarvianSpekkens2013} I. Marvian and R. W. Spekkens, New J. Phys. \textbf{15}, 033001 (2013).





\bibitem {Jozsa-Schlienz}R. Jozsa and J. Schlienz, Phys.Rev. A \textbf{62},
012301-1 (1999).

\bibitem{Caves} C. M. Caves, Phys. Rev. D \textbf{26}, 1817 (1982).

\bibitem{Caves2} C. M. Caves, J. Combes, Z. Jiang, S. Pandey, Phys. Rev. A {\bf 86}, 063802 (2012); S. Pandey, Z. Jiang, J. Combes, C. M. Caves, arXiv:1304.3901v2 (quant-ph).

\bibitem{Kitaev-Mayers-Preskill} A. Kitaev, D. Mayers, and J. Preskill, Phys. Rev. A \textbf{69}, 052326 (2004).


\bibitem{Hayashi} M. Hayashi, \textit{Quantum Information: An Introduction}, (Springer, 2006).

\bibitem{Helstrom} C. W. Helstrom, \textit{Quantum detection and estimation theory}, (Academic press,1976).

\bibitem{Aberg} J. \r{A}berg, arXiv:quant-ph/0612146; T.~Baumgratz, M.~Cramer, M.~B.~Plenio, arXiv:1311.0275 (quant-ph).



\bibitem{GourMarvianSpekkens} G. Gour, I. Marvian and R. W. Spekkens, Phys. Rev. A {\bf 80}, 012307 (2009).

\bibitem{MarvianSpekkensModes} I. Marvian and R. W. Spekkens, unpublished.

\bibitem{Luo} S. Luo, Phys. Rev. Lett. {\bf 91}, 180403 (2003).

\bibitem{Nielsen} M. A. Nielsen, and I. L. Chuang, \textit{Quantum Computation and Quantum Information}, (Cambridge University Press, Cambridge, 2000).

\end{thebibliography}
\end{document}